\documentclass[reprint,amsmath,amssymb,aps,prl,twocolumn,showpacs,superscriptaddress]{revtex4-1}

\usepackage{graphicx}	
\begin{document}

\title{Direct Optical Coupling to an Unoccupied Dirac Surface State in the Topological Insulator Bi$_2$Se$_3$}

\author{J.~A. Sobota}
\affiliation{Stanford Institute for Materials and Energy Sciences, SLAC National Accelerator Laboratory, 2575 Sand Hill Road, Menlo Park, CA 94025, USA}
\affiliation{Geballe Laboratory for Advanced Materials, Department of Applied Physics, Stanford University, Stanford, CA 94305, USA}
\affiliation{Department of Physics, Stanford University, Stanford, CA 94305, USA}
\author{S.-L. Yang}
\affiliation{Stanford Institute for Materials and Energy Sciences, SLAC National Accelerator Laboratory, 2575 Sand Hill Road, Menlo Park, CA 94025, USA}
\affiliation{Geballe Laboratory for Advanced Materials, Department of Applied Physics, Stanford University, Stanford, CA 94305, USA}
\affiliation{Department of Physics, Stanford University, Stanford, CA 94305, USA}
\author{A.~F. Kemper}
\affiliation{Lawrence Berkeley National Lab, 1 Cyclotron Road, Berkeley, CA 94720}
\author{J.~J. Lee}
\affiliation{Stanford Institute for Materials and Energy Sciences, SLAC National Accelerator Laboratory, 2575 Sand Hill Road, Menlo Park, CA 94025, USA}
\affiliation{Geballe Laboratory for Advanced Materials, Department of Applied Physics, Stanford University, Stanford, CA 94305, USA}
\affiliation{Department of Physics, Stanford University, Stanford, CA 94305, USA}
\author{F.~T. Schmitt}
\author{W. Li}
\author{R.~G. Moore}
\affiliation{Stanford Institute for Materials and Energy Sciences, SLAC National Accelerator Laboratory, 2575 Sand Hill Road, Menlo Park, CA 94025, USA}
\affiliation{Geballe Laboratory for Advanced Materials, Department of Applied Physics, Stanford University, Stanford, CA 94305, USA}
\author{J.~G. Analytis}
\affiliation{Department of Physics, University of California, Berkeley, California 94720, USA}
\author{I.~R. Fisher}
\affiliation{Stanford Institute for Materials and Energy Sciences, SLAC National Accelerator Laboratory, 2575 Sand Hill Road, Menlo Park, CA 94025, USA}
\affiliation{Geballe Laboratory for Advanced Materials, Department of Applied Physics, Stanford University, Stanford, CA 94305, USA}
\author{P.~S. Kirchmann}
\email{kirchman@stanford.edu}
\affiliation{Stanford Institute for Materials and Energy Sciences, SLAC National Accelerator Laboratory, 2575 Sand Hill Road, Menlo Park, CA 94025, USA}
\author{T.~P. Devereaux}
\affiliation{Stanford Institute for Materials and Energy Sciences, SLAC National Accelerator Laboratory, 2575 Sand Hill Road, Menlo Park, CA 94025, USA}
\affiliation{Geballe Laboratory for Advanced Materials, Department of Applied Physics, Stanford University, Stanford, CA 94305, USA}
\author{Z.-X. Shen}
\email{zxshen@stanford.edu}
\affiliation{Stanford Institute for Materials and Energy Sciences, SLAC National Accelerator Laboratory, 2575 Sand Hill Road, Menlo Park, CA 94025, USA}
\affiliation{Geballe Laboratory for Advanced Materials, Department of Applied Physics, Stanford University, Stanford, CA 94305, USA}
\affiliation{Department of Physics, Stanford University, Stanford, CA 94305, USA}

\date{\today}

\begin{abstract}

We characterize the occupied and unoccupied electronic structure of the topological insulator Bi$_2$Se$_3$ by  one-photon and two-photon angle-resolved photoemission spectroscopy and slab band structure calculations. We reveal a second, unoccupied Dirac surface state with similar electronic structure and physical origin to the well-known topological surface state. This state is energetically located 1.5~eV above the conduction band, which permits it to be directly excited by the output of a Ti:Sapphire laser. This discovery demonstrates the feasibility of direct ultrafast optical coupling to a topologically protected, spin-textured surface state.

\end{abstract}

\maketitle


Three dimensional topological insulators (TIs) are materials characterized by an insulating bulk and a conductive surface electronic structure. The hallmark of a TI is its linearly dispersing surface state (SS) which is guaranteed to cross the band gap separating the bulk valence band (VB) and conduction band (CB) \cite{Fu2007,Zhang2009,Chen2009,Xia2009}. In addition to this so-called topological protection, the SS has a chiral spin texture \cite{Hsieh2009a,Hsieh2009b} which offers the electrons protection against backscattering and has great appeal for spintronics applications \cite{Roushan2009,Garate2010}. A number of interesting phenomena are associated with optical coupling to TIs, such as colossal Kerr rotation \cite{Aguilar2011}, divergent photon absorption \cite{Wang2012}, spin transport \cite{Hosur2011,McIver2012}, and a long-lived SS population \cite{Sobota2012}. To exploit these phenomena, a detailed understanding of the response to optical excitation is required. A number of studies have investigated  electron and phonon dynamics initiated by ultrafast optical excitation either by optical reflectivity \cite{Qi2010,Kumar2011}, second harmonic generation \cite{Hsieh2011a}, or time- and angle- resolved photoemission spectroscopy (trARPES) \cite{Sobota2012,Wang2012a,Hajlaoui2012,Crepaldi2012}. However, these studies have not explicitly addressed the electronic transition driven by the excitation. A detailed understanding of the unoccupied electronic structure is required to understand precisely how photons couple to  electrons in these materials. Inverse photoemission has previously been applied to Bi$_2$Se$_3$, but the energy resolution of 0.56~eV and lack of momentum information precluded the ability to get a detailed picture of the unoccupied states \cite{Ueda1999}. More recently, significant progress was made by Niesner \emph{et. al.}, who used two-photon photoemission spectroscopy to investigate the material family Bi$_2$Te$_x$Se$_{3-x}$ \cite{Niesner2012}. Remarkably, they identified signatures of a second SS in the unoccupied states which is expected to have the same topological protection and chiral spin texture \cite{Niesner2012,Eremeev2013}. 

In this work, we use a combination of one-photon photoemission (1PPE) and two-photon photoemission (2PPE) to resolve the occupied and unoccupied electronic structure of $p-$ and $n-$type Bi$_2$Se$_3$, which is substantiated with theoretical band structure calculations. The energy and momentum resolution is drastically improved over previous work \cite{Niesner2012}, allowing an unambiguous identification of the second SS. Moreover, we show that in $n$-type samples, 1.5~eV photons drive a direct transition precisely into this state, thereby laying the foundation for direct ultrafast optical coupling to topological SSs. 

\begin{figure}
\resizebox{\columnwidth}{!}{\includegraphics{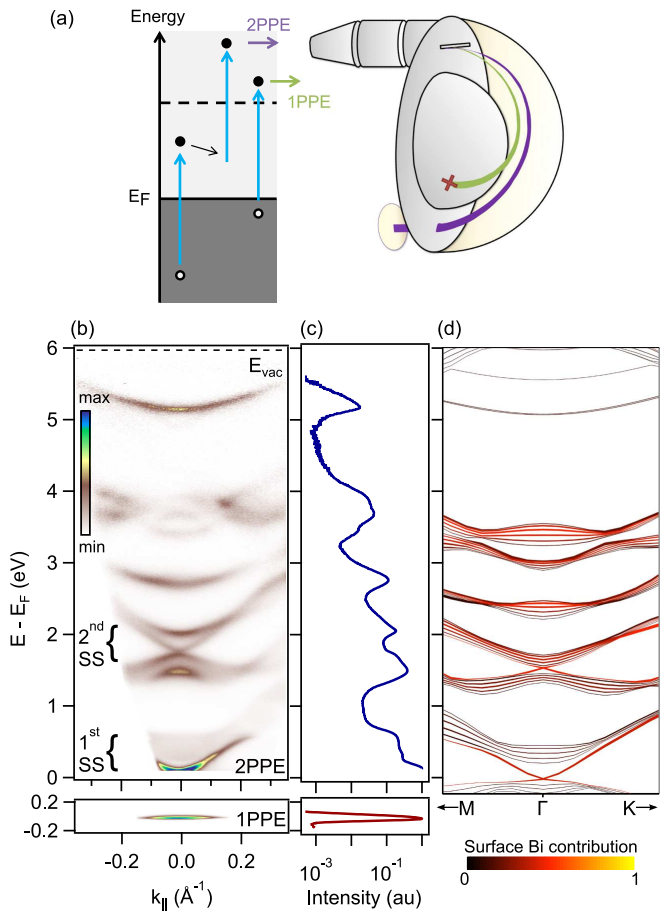}}
\caption{(Color online) (a) Schematic of the 1PPE and 2PPE processes utilized in this work. (b) 2PPE and 1PPE spectra of $p$-type Bi$_2$Se$_3$. The 1PPE spectrum contains only a narrow band of intensity from the occupied portion of the VB above the low-energy cutoff. The 2PPE spectrum reveals states from $E_F$ up to $E_{\textrm{vac}}$, including the well-known Dirac state around $\sim0.5$~eV, as well as a second Dirac state around $\sim1.8$~eV. The intensities are rescaled exponentially as a function of energy to make all features visible on the same image. (c) Momentum-integrated energy distribution curves showing the dynamic range of the 2PPE intensity. (d) Slab band structure calculation, which shows a one-to-one correspondence with the observed features.
\label{fig1}}
\end{figure}


Single crystals of $p$-type Bi$_2$Se$_3$ were synthesized as described in Ref. \onlinecite{Sobota2012}. These samples were cleaved \emph{in-situ} prior to measurement at a pressure $<5\times10^{-11}$ torr. Thin film $n$-type Bi$_2$Se$_3$ was grown on a sapphire substrate by molecular beam epitaxy as described in Ref. \onlinecite{Lee2012} and transported while under vacuum to our experimental chamber for measurement. Photoemission was performed with linearly polarized 5.98~eV and 1.5~eV photons derived from a Ti:Sapphire oscillator operating at a repetition rate of 80~MHz. The photoelectron kinetic energy and emission angle were resolved by a hemispherical electron analyzer. The energy resolution was determined to be 16~meV. All measurements were performed with the sample at low temperature (between 10 and 20K). Additional details on the experimental setup are given in Ref. \onlinecite{Sobota2012}. 

The density functional theory calculations were done with the PBE exchange functional \cite{Perdew1996}
using the full potential (linearized) augmented plane-wave method
as implemented in the \textsc{wien2k} package\cite{Blaha2001}. The calculations were based on
the experimentally determined crystal structure reported in Ref. \onlinecite{Nakajima1963}. 
The Bi$_2$Se$_3$ slab consists of 6 quintuple
layers, separated by a vacuum of $\sim 25$\AA.
We used a $k$-mesh size of $15\times15\times1$, and utilized spin-orbit coupling
in the self-consistent calculations unless otherwise indicated.


Unless otherwise specified, the measurements in this work were performed with 5.98~eV photons via 1PPE or 2PPE processes. In a 1PPE process, which is the basis for conventional ARPES measurements, a single photon promotes an electron from below the Fermi level $E_F$ to above the vacuum level $E_{\textrm{vac}}$. It is thereby emitted from the sample surface and collected by a photoelectron analyzer \cite{Hufner1995}. In 2PPE a photon first promotes an electron from below $E_F$ to an unoccupied intermediate state, and a second photon subsequently excites the electron above $E_{\textrm{vac}}$ \cite{Haight1995,Petek1998,Weinelt2002,Ohtsubo2012}. Due to the finite time duration of the laser pulses (of order 100~fs in our case), electron relaxation processes can occur in the intermediate states before the electron is photoemitted. These processes are illustrated in Fig. 1(a). Conceptually, the primary distinction is that 1PPE can measure only occupied states, while 2PPE grants access to unoccupied states. Our experimental configuration is identical for both measurements, except for the incident photon intensity. Since 1PPE scales linearly with peak intensity and 2PPE scales quadratically, we switch between 1PPE and 2PPE acquisition modes merely by tuning the intensity. We stress that this is a significant distinction between our measurement and standard 2PPE measurements:  Conventional 2PPE is performed with the photon energies chosen to be less than the sample work function, so that an overwhelmingly intense 1PPE signal is avoided. Here we deliberately use photon energies larger than the work function, allowing us to perform both 1PPE and 2PPE. To avoid the intense 1PPE signal while measuring in 2PPE mode, we operate the analyzer such that only electrons with energy $>E_F$ are collected, as illustrated in the cartoon of Fig. 1(a).

\begin{figure}
\resizebox{\columnwidth}{!}{\includegraphics{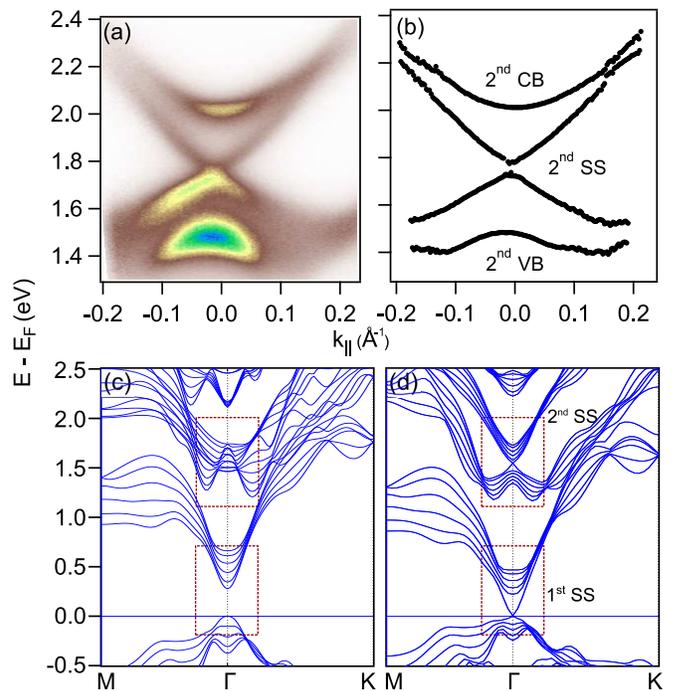}}
\caption{(Color online) (a) 2PPE spectrum of the unoccupied states and (b) peak positions obtained from EDC fitting. The unoccupied Dirac cone is unambiguously resolved and crosses the gap between two bulk states. (c) and (d) Theoretical band structure calculations without and with spin-orbit coupling, respectively. Both Dirac cones only exist with SOI, suggesting a similar physical origin for both states.
\label{fig2}}
\end{figure}

We begin by presenting the 2PPE and 1PPE spectra of $p$-type Bi$_2$Se$_3$ in the upper and lower panels, respectively, of Fig. 1(b). The 1PPE spectrum consists of only a narrow band of intensity from the occupied portion of the VB above the low-energy cutoff. The reason for this narrow band is that the work function of 5.97~eV is barely exceeded by the photon energy of 5.98~eV. Due to the $p$-type doping, the SS and CB are above $E_F$, and are absent in the 1PPE spectrum \cite{Sobota2012}.

To identify the more novel spectral features, we now turn to the 2PPE spectrum, which reveals states all the way from $E_F$ up to $E_{\textrm{vac}}$. Here we observe the familiar CB and Dirac SS between  $E-E_F=0$ and 0.7~eV. These states have the usual characteristics observed in conventional ARPES measurements of $n$-type materials: There is a bandgap of $\sim$0.2~eV which is crossed by the linearly dispersive SS \cite{Chen2009,Xia2009,Chen2010}. We now focus our attention on states which lie at higher energy. We note that the spectral intensity spans a dynamic range of $\sim$3 orders of magnitude (see Fig.1(c)) and so the intensities of Fig. 1(b) are rescaled exponentially with energy to allow all features to be displayed simultaneously.

A multitude of features are present in the 2PPE spectrum. In principle, the 2PPE spectral intensity can be complicated by contributions from initial and final states, in addition to the intermediate states which we wish to study \cite{Petek1998,Weinelt2002}. To help clarify the origin of the observed features, we display the calculated band structure in Fig. 1(d). While the energy scales are not reproduced exactly, there is an unambiguous one-to-one correspondence between the observed features and those appearing in the calculation. This gives us confidence that the 2PPE spectrum is representative of the intermediate states.

The most intriguing spectral features exist between 1.3 and 2.4~eV, and are shown in detail in Fig. 2(a). Four dispersive bands are observed in this window; for greater clarity, the peak positions obtained by energy distribution curve (EDC) fitting are shown in panel (b). The band structure is strikingly similar to the VB, CB, and SS bands near $E_F$. In fact, Niesner \emph{et. al.} identified the linearly dispersive band as an unoccupied topologically protected, spin chiral SS, while the upper and lower bands are the corresponding bulk bands\cite{Niesner2012}. To distinguish these two sets of bands, we refer to the Dirac cone near $E_F$ as the $1^\textrm{st}$ SS, and the higher energy Dirac cone as the $2^\textrm{nd}$ SS. The corresponding bulk bands are referred to as the $1^\textrm{st}$ and $2^\textrm{nd}$ VB and CB. We note that the  experimental resolution in the previous work did not allow for an unambiguous identification of the $2^\textrm{nd}$ SS in the gap between the bulk states. Here the spectral separation between the four bands is clearly resolved, allowing us to study the band structure in greater detail. In particular, we resolve the two bulk bands separated by a gap of 530~meV, which is crossed by the linearly dispersive SS. Note, however, that this estimate for the band gap must be regarded as an upper bound for the true value, since it is not known whether the observed electrons are probed from $k_z = 0$ \cite{Hufner1995}. The $2^\textrm{nd}$ SS disperses with a velocity of  $3.3\times10^5$~m/s at the Dirac point, which is comparable to the velocity of $5.4\times10^5$~m/s typically measured for the Fermi velocity of the $1^\textrm{st}$ SS \cite{Kuroda2010}. Our calculations confirm the assignment of these bands to bulk and surface states (Fig. 1(d)). Moreover, just like the $1^\textrm{st}$ SS, the $2^\textrm{nd}$ SS exists only in the presence of crystal spin-orbit coupling, as evidenced by the calculations in Fig. 2(c) and (d). This suggests that both SSs share the same physical origin, as they both arise due to symmetry inversion of bulk states in the presence of strong spin-orbit coupling.

To better characterize the $2^\textrm{nd}$ SS, we performed a band mapping to determine its 2-dimensional dispersion. Since band mapping entails rotation of the sample relative to the analyzer, it benefits greatly from having a flat and homogeneous sample. For this we used Bi$_2$Se$_3$ thin films grown by molecular beam epitaxy. These films have $n$-type doping, but we have confirmed that the band structure, including the $1^\textrm{st}$ and $2^\textrm{nd}$ SSs, is equivalent to that of the $p$-type crystals. Most importantly, they are extremely flat due to the sapphire substrate upon which they were grown \cite{Lee2012}.

The results of the band mapping are summarized in the constant energy maps of Fig. 3. All energies are referenced to the Dirac point $E_D$ of the $2^\textrm{nd}$ SS, as shown in panel (a). These results demonstrate that the band structure evolves from a hexagonal bulk band (b), to a circular SS (c-e), to a hexagonal bulk band (f). This is illustrated schematically in Fig. 3(g). This is reminiscent of the $1^\textrm{st}$ SS, which is known to have a circular geometry and be situated between the hexagonal VB and CB \cite{Chen2010,Kuroda2010}. The similarity of their band geometries provides further evidence that the $1^\textrm{st}$ and $2^\textrm{nd}$ SSs have similar physical origins.

\begin{figure}
\resizebox{\columnwidth}{!}{\includegraphics{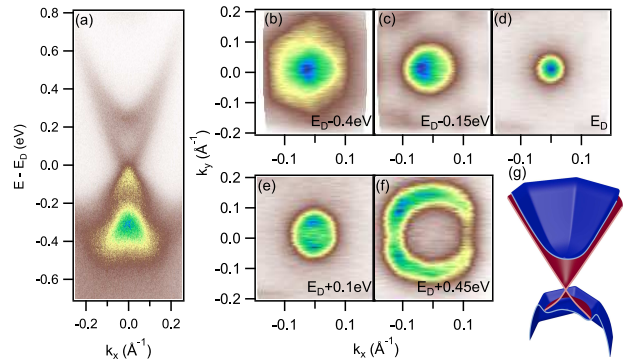}}
\caption{(Color online)   (a) 6~eV 2PPE spectrum of the $2^\textrm{nd}$ SS on an $n$-type Bi$_2$Se$_3$ thin film. (b)-(f) Constant energy maps. The electronic structure evolves from a hexagonal VB, to a circular SS, to a hexagonal CB. Energies are referenced to the Dirac point $E_D$. (g) Schematic of the band geometry.
\label{fig3}}
\end{figure}

\begin{figure}
\resizebox{\columnwidth}{!}{\includegraphics{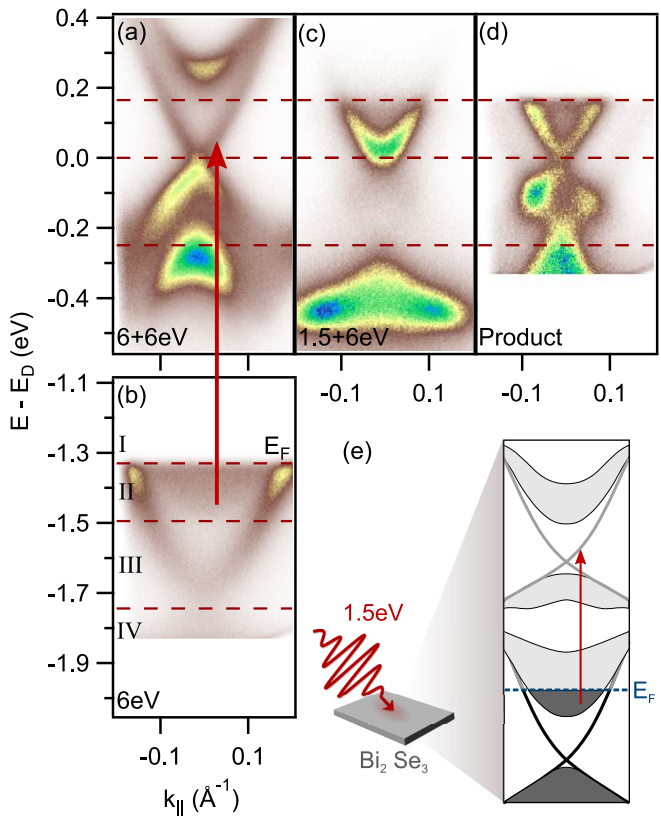}}
\caption{(Color online) (a) 2PPE spectrum of $2^\textrm{nd}$ SS using 6~eV photons on the $p$-type sample. (b) ARPES spectrum of $n$-type Bi$_2$Se$_3$ in the vicinity of $E_F$. Both the $1^\textrm{st}$ SS and CB are occupied. Note that the energy axis is referenced to the Dirac point of the $2^\textrm{nd}$ SS. (c) 2PPE spectrum using 1.5~eV photons to populate the $2^\textrm{nd}$ SS, and 6~eV photons to photoemit. (d) Multiplication of the spectra in (a) and (b), showing that the 2PPE spectrum in (c) can be understood as a projection of initial states onto intermediate states. (e) Schematic illustration of the 1.5~eV photoexcitation process in Bi$_2$Se$_3$.
\label{fig4}}
\end{figure}

We now discuss the relevance of the $2^\textrm{nd}$ SS to studies and applications utilizing ultrafast optical excitation. These applications typically use a photon energy of 1.5~eV due to the use of a Ti:Sapphire laser source, so we investigate the use of 1.5~eV photons to couple to the $2^\textrm{nd}$ SS. In Fig. 4(a) we again show the unoccupied $2^\textrm{nd}$ SS obtained by 6~eV 2PPE, and in (b) we show the occupied bandstructure of the $n$-type sample obtained by 6~eV 1PPE. Coincidentally, we see that the $2^\textrm{nd}$ Dirac point is located almost precisely 1.5~eV above the $1^\textrm{st}$ CB edge, which suggests the possibility of direct ultrafast optical excitation of the  $2^\textrm{nd}$ SS by driving a transition from the $1^\textrm{st}$ CB. We demonstrate the feasibility of this approach in Fig. 4(c), which is a 2PPE spectrum in which 1.5~eV photons are used to transiently populate the unoccupied states, and 6~eV photons to photoemit. No electrons are excited to states above $E-E_D$ = 0.17~eV since the corresponding initial states lie above $E_F$  (region I in panel (b)). However, as we expected, the 1.5~eV excitation causes the $2^\textrm{nd}$ SS to get populated with electrons from between $E_F$ and the CB edge (region II). The states between -0.24 and 0~eV get only weakly populated, reflecting the fact that few initial states are available due to the bulk bandgap (region III). However, a weak population is observed due to electrons being excited from the $1^\textrm{st}$ SS. The population below -0.24~eV  is attributable to a transition from the $1^\textrm{st}$ VB (region IV	) to the $2^\textrm{nd}$ VB. 

The spectrum of Fig. 4(c) can be understood intuitively as the projection of the initial states below $E_F$ to the unoccupied intermediate states. Fig. 4(a) and (b) can be taken as approximations to the intermediate and initial state band structures, respectively, and we  multiply these two spectra as a way of approximating the projection. The result, shown in Fig. 4(d), indeed bears a strong resemblance to the actual 2PPE spectrum in (c). The product spectrum correctly shows the population of the $2^\textrm{nd}$ SS attributable to the $1^\textrm{st}$ CB in region II, as well as the population in region IV due to the transition between the two VBs. The only shortcoming is that it overstates the population coming from the  $1^\textrm{st}$ SS in region III. However, since this discussion has neglected matrix elements and final state effects, it provides only a qualitative picture. A cartoon illustration of 1.5~eV photoexcitation of Bi$_2$Se$_3$  is provided in Fig. 4(e).

In summary, we have characterized the electronic structure of Bi$_2$Se$_3$ with a combination of 1PPE and 2PPE techniques. 1PPE with 6~eV photons probes the electronic structure below $E_F$, as is typical for conventional ARPES measurements, while 2PPE with 6~eV photons illuminates the states between $E_F$ and $E_{\textrm{vac}}$. This technique reveals a $2^\textrm{nd}$ Dirac SS located 1.5~eV above the $1^\textrm{st}$ CB edge, and we have further shown that excitation with 1.5~eV photons drives a direct transition between these two states. 

There are a number of interesting implications of these results. First, to our knowledge, this is the first measurement on any material  using a single photon energy for both 1PPE and 2PPE. This is achieved by using a photon energy exceeding the sample work function, which has conventionally been avoided in 2PPE measurements. Since the only experimental parameter  tuned between 1PPE and 2PPE modes of operation is the intensity, this represents an exceptionally simple method to  determine both the occupied and unoccupied band structure of a material. This has exciting implications, for example, for materials such as the high-$T_c$ cuprates, where a measurement of both the occupied and unoccupied sides of the energy gap would provide insight on the particle-hole asymmetry of the superconducting and pseudogap states \cite{Hashimoto2010,Moritz2011}.

Next, we emphasize how this work builds on that of Niesner \emph{et. al.}  by using superior spectral resolution to provide unambiguous evidence of the $2^\textrm{nd}$ SS \cite{Niesner2012}. This state is expected to have all the novel topological and spin properties which characterize the well-known $1^\textrm{st}$ Dirac SS of TIs \cite{Niesner2012,Eremeev2013}. Our demonstration that it is populated by 1.5~eV photons could have relevance for the number of existing studies on $n$-type TIs which have utilized 1.5~eV photons \cite{Qi2010,Kumar2011,Hsieh2011a,Wang2012a,Hajlaoui2012,Crepaldi2012,McIver2012}. These experiments have been interpreted without knowledge of the unoccupied topological SS, and it would be interesting to evaluate whether the transition into this state plays a role in the physics discussed in those results. Finally, the fact that it can be accessed by 1.5~eV photons is particularly appropriate for applications, since this is the fundamental photon energy provided by commercial ultrafast Ti:Sapphire lasers. This discovery therefore demonstrates a unique opportunity for direct ultrafast optical coupling to TI SSs.

\begin{acknowledgments}
This work is supported by the Department of Energy, Office of Basic Energy Sciences, Division of Materials Science. JAS acknowledges support by the Stanford Graduate Fellowship. AFK is supported by the Laboratory Directed Research and Development Program of Lawrence Berkeley National Laboratory under the U.S. Department of Energy contract number DE-AC02-05CH11231. 
\end{acknowledgments}

\bibliography{2PPE_BIB}

\begin{thebibliography}{35}%
\makeatletter
\providecommand \@ifxundefined [1]{%
 \@ifx{#1\undefined}
}%
\providecommand \@ifnum [1]{%
 \ifnum #1\expandafter \@firstoftwo
 \else \expandafter \@secondoftwo
 \fi
}%
\providecommand \@ifx [1]{%
 \ifx #1\expandafter \@firstoftwo
 \else \expandafter \@secondoftwo
 \fi
}%
\providecommand \natexlab [1]{#1}%
\providecommand \enquote  [1]{``#1''}%
\providecommand \bibnamefont  [1]{#1}%
\providecommand \bibfnamefont [1]{#1}%
\providecommand \citenamefont [1]{#1}%
\providecommand \href@noop [0]{\@secondoftwo}%
\providecommand \href [0]{\begingroup \@sanitize@url \@href}%
\providecommand \@href[1]{\@@startlink{#1}\@@href}%
\providecommand \@@href[1]{\endgroup#1\@@endlink}%
\providecommand \@sanitize@url [0]{\catcode `\\12\catcode `\$12\catcode
  `\&12\catcode `\#12\catcode `\^12\catcode `\_12\catcode `\%12\relax}%
\providecommand \@@startlink[1]{}%
\providecommand \@@endlink[0]{}%
\providecommand \url  [0]{\begingroup\@sanitize@url \@url }%
\providecommand \@url [1]{\endgroup\@href {#1}{\urlprefix }}%
\providecommand \urlprefix  [0]{URL }%
\providecommand \Eprint [0]{\href }%
\providecommand \doibase [0]{http://dx.doi.org/}%
\providecommand \selectlanguage [0]{\@gobble}%
\providecommand \bibinfo  [0]{\@secondoftwo}%
\providecommand \bibfield  [0]{\@secondoftwo}%
\providecommand \translation [1]{[#1]}%
\providecommand \BibitemOpen [0]{}%
\providecommand \bibitemStop [0]{}%
\providecommand \bibitemNoStop [0]{.\EOS\space}%
\providecommand \EOS [0]{\spacefactor3000\relax}%
\providecommand \BibitemShut  [1]{\csname bibitem#1\endcsname}%
\let\auto@bib@innerbib\@empty
\bibitem [{\citenamefont {Fu}\ \emph {et~al.}(2007)\citenamefont {Fu},
  \citenamefont {Kane},\ and\ \citenamefont {Mele}}]{Fu2007}%
  \BibitemOpen
  \bibfield  {author} {\bibinfo {author} {\bibfnamefont {L.}~\bibnamefont
  {Fu}}, \bibinfo {author} {\bibfnamefont {C.}~\bibnamefont {Kane}}, \ and\
  \bibinfo {author} {\bibfnamefont {E.}~\bibnamefont {Mele}},\ }\href@noop {}
  {\bibfield  {journal} {\bibinfo  {journal} {Phys. Rev. Lett.}\ }\textbf
  {\bibinfo {volume} {98}},\ \bibinfo {pages} {106803} (\bibinfo {year}
  {2007})}\BibitemShut {NoStop}%
\bibitem [{\citenamefont {Zhang}\ \emph {et~al.}(2009)\citenamefont {Zhang}
  \emph {et~al.}}]{Zhang2009}%
  \BibitemOpen
  \bibfield  {author} {\bibinfo {author} {\bibfnamefont {H.}~\bibnamefont
  {Zhang}} \emph {et~al.},\ }\href@noop {} {\bibfield  {journal} {\bibinfo
  {journal} {Nature Phys.}\ }\textbf {\bibinfo {volume} {5}},\ \bibinfo {pages}
  {438} (\bibinfo {year} {2009})}\BibitemShut {NoStop}%
\bibitem [{\citenamefont {Chen}\ \emph {et~al.}(2009)\citenamefont {Chen} \emph
  {et~al.}}]{Chen2009}%
  \BibitemOpen
  \bibfield  {author} {\bibinfo {author} {\bibfnamefont {Y.~L.}\ \bibnamefont
  {Chen}} \emph {et~al.},\ }\href@noop {} {\bibfield  {journal} {\bibinfo
  {journal} {Science}\ }\textbf {\bibinfo {volume} {325}},\ \bibinfo {pages}
  {178} (\bibinfo {year} {2009})}\BibitemShut {NoStop}%
\bibitem [{\citenamefont {Xia}\ \emph {et~al.}(2009)\citenamefont {Xia} \emph
  {et~al.}}]{Xia2009}%
  \BibitemOpen
  \bibfield  {author} {\bibinfo {author} {\bibfnamefont {Y.}~\bibnamefont
  {Xia}} \emph {et~al.},\ }\href@noop {} {\bibfield  {journal} {\bibinfo
  {journal} {Nature Phys.}\ }\textbf {\bibinfo {volume} {5}},\ \bibinfo {pages}
  {398} (\bibinfo {year} {2009})}\BibitemShut {NoStop}%
\bibitem [{\citenamefont {Hsieh}\ \emph
  {et~al.}(2009{\natexlab{a}})\citenamefont {Hsieh} \emph
  {et~al.}}]{Hsieh2009a}%
  \BibitemOpen
  \bibfield  {author} {\bibinfo {author} {\bibfnamefont {D.}~\bibnamefont
  {Hsieh}} \emph {et~al.},\ }\href@noop {} {\bibfield  {journal} {\bibinfo
  {journal} {Science}\ }\textbf {\bibinfo {volume} {323}},\ \bibinfo {pages}
  {919} (\bibinfo {year} {2009}{\natexlab{a}})}\BibitemShut {NoStop}%
\bibitem [{\citenamefont {Hsieh}\ \emph
  {et~al.}(2009{\natexlab{b}})\citenamefont {Hsieh} \emph
  {et~al.}}]{Hsieh2009b}%
  \BibitemOpen
  \bibfield  {author} {\bibinfo {author} {\bibfnamefont {D.}~\bibnamefont
  {Hsieh}} \emph {et~al.},\ }\href@noop {} {\bibfield  {journal} {\bibinfo
  {journal} {Nature}\ }\textbf {\bibinfo {volume} {460}},\ \bibinfo {pages}
  {1101} (\bibinfo {year} {2009}{\natexlab{b}})}\BibitemShut {NoStop}%
\bibitem [{\citenamefont {Roushan}\ \emph {et~al.}(2009)\citenamefont {Roushan}
  \emph {et~al.}}]{Roushan2009}%
  \BibitemOpen
  \bibfield  {author} {\bibinfo {author} {\bibfnamefont {P.}~\bibnamefont
  {Roushan}} \emph {et~al.},\ }\href@noop {} {\bibfield  {journal} {\bibinfo
  {journal} {Nature}\ }\textbf {\bibinfo {volume} {460}},\ \bibinfo {pages}
  {1106} (\bibinfo {year} {2009})}\BibitemShut {NoStop}%
\bibitem [{\citenamefont {Garate}\ and\ \citenamefont
  {Franz}(2010)}]{Garate2010}%
  \BibitemOpen
  \bibfield  {author} {\bibinfo {author} {\bibfnamefont {I.}~\bibnamefont
  {Garate}}\ and\ \bibinfo {author} {\bibfnamefont {M.}~\bibnamefont {Franz}},\
  }\href@noop {} {\bibfield  {journal} {\bibinfo  {journal} {Phys. Rev. Lett.}\
  }\textbf {\bibinfo {volume} {104}},\ \bibinfo {pages} {146802} (\bibinfo
  {year} {2010})}\BibitemShut {NoStop}%
\bibitem [{\citenamefont {Aguilar}\ \emph {et~al.}(2011)\citenamefont {Aguilar}
  \emph {et~al.}}]{Aguilar2011}%
  \BibitemOpen
  \bibfield  {author} {\bibinfo {author} {\bibfnamefont {R.~V.}\ \bibnamefont
  {Aguilar}} \emph {et~al.},\ }\href {http://arxiv.org/abs/1105.0237} {}
  (\bibinfo {year} {2011}),\ \Eprint {http://arxiv.org/abs/1105.0237}
  {arXiv:1105.0237} \BibitemShut {NoStop}%
\bibitem [{\citenamefont {Wang}\ \emph
  {et~al.}(2012{\natexlab{a}})\citenamefont {Wang}, \citenamefont {Mabuchi},\
  and\ \citenamefont {Qi}}]{Wang2012}%
  \BibitemOpen
  \bibfield  {author} {\bibinfo {author} {\bibfnamefont {J.}~\bibnamefont
  {Wang}}, \bibinfo {author} {\bibfnamefont {H.}~\bibnamefont {Mabuchi}}, \
  and\ \bibinfo {author} {\bibfnamefont {X.-L.}\ \bibnamefont {Qi}},\ }\href
  {http://arxiv.org/abs/1209.6597} {} (\bibinfo {year} {2012}{\natexlab{a}}),\
  \Eprint {http://arxiv.org/abs/1209.6597} {arXiv:1209.6597} \BibitemShut
  {NoStop}%
\bibitem [{\citenamefont {Hosur}(2011)}]{Hosur2011}%
  \BibitemOpen
  \bibfield  {author} {\bibinfo {author} {\bibfnamefont {P.}~\bibnamefont
  {Hosur}},\ }\href@noop {} {\bibfield  {journal} {\bibinfo  {journal} {Phys.
  Rev. B}\ }\textbf {\bibinfo {volume} {83}},\ \bibinfo {pages} {035309}
  (\bibinfo {year} {2011})}\BibitemShut {NoStop}%
\bibitem [{\citenamefont {McIver}\ \emph {et~al.}(2012)\citenamefont {McIver}
  \emph {et~al.}}]{McIver2012}%
  \BibitemOpen
  \bibfield  {author} {\bibinfo {author} {\bibfnamefont {J.~W.}\ \bibnamefont
  {McIver}} \emph {et~al.},\ }\href@noop {} {\bibfield  {journal} {\bibinfo
  {journal} {Nature Nanotech.}\ }\textbf {\bibinfo {volume} {7}},\ \bibinfo
  {pages} {96} (\bibinfo {year} {2012})}\BibitemShut {NoStop}%
\bibitem [{\citenamefont {Sobota}\ \emph {et~al.}(2012)\citenamefont {Sobota}
  \emph {et~al.}}]{Sobota2012}%
  \BibitemOpen
  \bibfield  {author} {\bibinfo {author} {\bibfnamefont {J.~A.}\ \bibnamefont
  {Sobota}} \emph {et~al.},\ }\href@noop {} {\bibfield  {journal} {\bibinfo
  {journal} {Phys. Rev. Lett.}\ }\textbf {\bibinfo {volume} {108}},\ \bibinfo
  {pages} {117403} (\bibinfo {year} {2012})}\BibitemShut {NoStop}%
\bibitem [{\citenamefont {Qi}\ \emph {et~al.}(2010)\citenamefont {Qi} \emph
  {et~al.}}]{Qi2010}%
  \BibitemOpen
  \bibfield  {author} {\bibinfo {author} {\bibfnamefont {J.}~\bibnamefont {Qi}}
  \emph {et~al.},\ }\href@noop {} {\bibfield  {journal} {\bibinfo  {journal}
  {Appl. Phys. Lett.}\ }\textbf {\bibinfo {volume} {97}},\ \bibinfo {pages}
  {182102} (\bibinfo {year} {2010})}\BibitemShut {NoStop}%
\bibitem [{\citenamefont {Kumar}\ \emph {et~al.}(2011)\citenamefont {Kumar}
  \emph {et~al.}}]{Kumar2011}%
  \BibitemOpen
  \bibfield  {author} {\bibinfo {author} {\bibfnamefont {N.}~\bibnamefont
  {Kumar}} \emph {et~al.},\ }\href@noop {} {\bibfield  {journal} {\bibinfo
  {journal} {Phys. Rev. B}\ }\textbf {\bibinfo {volume} {83}},\ \bibinfo
  {pages} {235306} (\bibinfo {year} {2011})}\BibitemShut {NoStop}%
\bibitem [{\citenamefont {Hsieh}\ \emph {et~al.}(2011)\citenamefont {Hsieh}
  \emph {et~al.}}]{Hsieh2011a}%
  \BibitemOpen
  \bibfield  {author} {\bibinfo {author} {\bibfnamefont {D.}~\bibnamefont
  {Hsieh}} \emph {et~al.},\ }\href@noop {} {\bibfield  {journal} {\bibinfo
  {journal} {Phys. Rev. Lett.}\ }\textbf {\bibinfo {volume} {107}},\ \bibinfo
  {pages} {077401} (\bibinfo {year} {2011})}\BibitemShut {NoStop}%
\bibitem [{\citenamefont {Wang}\ \emph
  {et~al.}(2012{\natexlab{b}})\citenamefont {Wang} \emph {et~al.}}]{Wang2012a}%
  \BibitemOpen
  \bibfield  {author} {\bibinfo {author} {\bibfnamefont {Y.~H.}\ \bibnamefont
  {Wang}} \emph {et~al.},\ }\href@noop {} {\bibfield  {journal} {\bibinfo
  {journal} {Phys. Rev. Lett.}\ }\textbf {\bibinfo {volume} {109}},\ \bibinfo
  {pages} {127401} (\bibinfo {year} {2012}{\natexlab{b}})}\BibitemShut
  {NoStop}%
\bibitem [{\citenamefont {Hajlaoui}\ \emph {et~al.}(2012)\citenamefont
  {Hajlaoui} \emph {et~al.}}]{Hajlaoui2012}%
  \BibitemOpen
  \bibfield  {author} {\bibinfo {author} {\bibfnamefont {M.}~\bibnamefont
  {Hajlaoui}} \emph {et~al.},\ }\href@noop {} {\bibfield  {journal} {\bibinfo
  {journal} {Nano Lett.}\ }\textbf {\bibinfo {volume} {12}},\ \bibinfo {pages}
  {3532} (\bibinfo {year} {2012})}\BibitemShut {NoStop}%
\bibitem [{\citenamefont {Crepaldi}\ \emph {et~al.}(2012)\citenamefont
  {Crepaldi} \emph {et~al.}}]{Crepaldi2012}%
  \BibitemOpen
  \bibfield  {author} {\bibinfo {author} {\bibfnamefont {A.}~\bibnamefont
  {Crepaldi}} \emph {et~al.},\ }\href@noop {} {\bibfield  {journal} {\bibinfo
  {journal} {Phys. Rev. B}\ }\textbf {\bibinfo {volume} {86}},\ \bibinfo
  {pages} {205133} (\bibinfo {year} {2012})}\BibitemShut {NoStop}%
\bibitem [{\citenamefont {Ueda}\ \emph {et~al.}(1999)\citenamefont {Ueda} \emph
  {et~al.}}]{Ueda1999}%
  \BibitemOpen
  \bibfield  {author} {\bibinfo {author} {\bibfnamefont {Y.}~\bibnamefont
  {Ueda}} \emph {et~al.},\ }\href@noop {} {\bibfield  {journal} {\bibinfo
  {journal} {J. Electron. Spectrosc. Relat. Phenom.}\ }\textbf {\bibinfo
  {volume} {101-103}},\ \bibinfo {pages} {677} (\bibinfo {year}
  {1999})}\BibitemShut {NoStop}%
\bibitem [{\citenamefont {Niesner}\ \emph {et~al.}(2012)\citenamefont {Niesner}
  \emph {et~al.}}]{Niesner2012}%
  \BibitemOpen
  \bibfield  {author} {\bibinfo {author} {\bibfnamefont {D.}~\bibnamefont
  {Niesner}} \emph {et~al.},\ }\href@noop {} {\bibfield  {journal} {\bibinfo
  {journal} {Phys. Rev. B}\ }\textbf {\bibinfo {volume} {86}},\ \bibinfo
  {pages} {205403} (\bibinfo {year} {2012})}\BibitemShut {NoStop}%
\bibitem [{\citenamefont {Eremeev}\ \emph {et~al.}(2013)\citenamefont {Eremeev}
  \emph {et~al.}}]{Eremeev2013}%
  \BibitemOpen
  \bibfield  {author} {\bibinfo {author} {\bibfnamefont {S.~V.}\ \bibnamefont
  {Eremeev}} \emph {et~al.},\ }\href@noop {} {\bibfield  {journal} {\bibinfo
  {journal} {JETP Lett.}\ }\textbf {\bibinfo {volume} {96}},\ \bibinfo {pages}
  {780} (\bibinfo {year} {2013})}\BibitemShut {NoStop}%
\bibitem [{\citenamefont {Lee}\ \emph {et~al.}(2012)\citenamefont {Lee} \emph
  {et~al.}}]{Lee2012}%
  \BibitemOpen
  \bibfield  {author} {\bibinfo {author} {\bibfnamefont {J.~J.}\ \bibnamefont
  {Lee}} \emph {et~al.},\ }\href@noop {} {\bibfield  {journal} {\bibinfo
  {journal} {Appl. Phys. Lett.}\ }\textbf {\bibinfo {volume} {101}},\ \bibinfo
  {pages} {013118} (\bibinfo {year} {2012})}\BibitemShut {NoStop}%
\bibitem [{\citenamefont {Perdew}\ \emph {et~al.}(1996)\citenamefont {Perdew},
  \citenamefont {Burke},\ and\ \citenamefont {Ernzerhof}}]{Perdew1996}%
  \BibitemOpen
  \bibfield  {author} {\bibinfo {author} {\bibfnamefont {J.~P.}\ \bibnamefont
  {Perdew}}, \bibinfo {author} {\bibfnamefont {K.}~\bibnamefont {Burke}}, \
  and\ \bibinfo {author} {\bibfnamefont {M.}~\bibnamefont {Ernzerhof}},\
  }\href@noop {} {\bibfield  {journal} {\bibinfo  {journal} {Phys. Rev. Lett.}\
  }\textbf {\bibinfo {volume} {77}},\ \bibinfo {pages} {3865} (\bibinfo {year}
  {1996})}\BibitemShut {NoStop}%
\bibitem [{\citenamefont {Blaha}\ \emph {et~al.}(2001)\citenamefont {Blaha}
  \emph {et~al.}}]{Blaha2001}%
  \BibitemOpen
  \bibfield  {author} {\bibinfo {author} {\bibfnamefont {P.}~\bibnamefont
  {Blaha}} \emph {et~al.},\ }\href@noop {} {\emph {\bibinfo {title}
  {{\textsc{wien2k}: An Augmented Plane Wave Plus Local Orbitals Program for
  Calculating Crystal Properties}}}},\ edited by\ \bibinfo {editor}
  {\bibfnamefont {K.}~\bibnamefont {Schwarz}}\ (\bibinfo  {publisher} {Techn.
  Universitat Wein, Austria},\ \bibinfo {year} {2001})\BibitemShut {NoStop}%
\bibitem [{\citenamefont {Nakajima}(1963)}]{Nakajima1963}%
  \BibitemOpen
  \bibfield  {author} {\bibinfo {author} {\bibfnamefont {S.}~\bibnamefont
  {Nakajima}},\ }\href@noop {} {\bibfield  {journal} {\bibinfo  {journal} {J.
  Phys. Chem. Solids}\ }\textbf {\bibinfo {volume} {24}},\ \bibinfo {pages}
  {479} (\bibinfo {year} {1963})}\BibitemShut {NoStop}%
\bibitem [{\citenamefont {H\"{u}fner}(1995)}]{Hufner1995}%
  \BibitemOpen
  \bibfield  {author} {\bibinfo {author} {\bibfnamefont {S.}~\bibnamefont
  {H\"{u}fner}},\ }\href@noop {} {\emph {\bibinfo {title} {{Photoelectron
  Spectroscopy}}}}\ (\bibinfo  {publisher} {Springer},\ \bibinfo {address}
  {Berlin},\ \bibinfo {year} {1995})\BibitemShut {NoStop}%
\bibitem [{\citenamefont {Haight}(1995)}]{Haight1995}%
  \BibitemOpen
  \bibfield  {author} {\bibinfo {author} {\bibfnamefont {R.}~\bibnamefont
  {Haight}},\ }\href@noop {} {\bibfield  {journal} {\bibinfo  {journal} {Surf.
  Sci. Rep.}\ }\textbf {\bibinfo {volume} {21}},\ \bibinfo {pages} {275}
  (\bibinfo {year} {1995})}\BibitemShut {NoStop}%
\bibitem [{Pet(1997)}]{Petek1998}%
  \BibitemOpen
  \href@noop {} {\bibfield  {journal} {\bibinfo  {journal} {Prog. in Surf.
  Sci.}\ }\textbf {\bibinfo {volume} {56}},\ \bibinfo {pages} {239 } (\bibinfo
  {year} {1997})}\BibitemShut {NoStop}%
\bibitem [{\citenamefont {Weinelt}(2002)}]{Weinelt2002}%
  \BibitemOpen
  \bibfield  {author} {\bibinfo {author} {\bibfnamefont {M.}~\bibnamefont
  {Weinelt}},\ }\href@noop {} {\bibfield  {journal} {\bibinfo  {journal} {J.
  Phys.: Condens. Matter}\ }\textbf {\bibinfo {volume} {14}},\ \bibinfo {pages}
  {R1099} (\bibinfo {year} {2002})}\BibitemShut {NoStop}%
\bibitem [{\citenamefont {Ohtsubo}\ \emph {et~al.}(2012)\citenamefont {Ohtsubo}
  \emph {et~al.}}]{Ohtsubo2012}%
  \BibitemOpen
  \bibfield  {author} {\bibinfo {author} {\bibfnamefont {Y.}~\bibnamefont
  {Ohtsubo}} \emph {et~al.},\ }\href@noop {} {\bibfield  {journal} {\bibinfo
  {journal} {Phys. Rev. Lett.}\ }\textbf {\bibinfo {volume} {109}},\ \bibinfo
  {pages} {226404} (\bibinfo {year} {2012})}\BibitemShut {NoStop}%
\bibitem [{\citenamefont {Chen}\ \emph {et~al.}(2010)\citenamefont {Chen} \emph
  {et~al.}}]{Chen2010}%
  \BibitemOpen
  \bibfield  {author} {\bibinfo {author} {\bibfnamefont {Y.~L.}\ \bibnamefont
  {Chen}} \emph {et~al.},\ }\href@noop {} {\bibfield  {journal} {\bibinfo
  {journal} {Science}\ }\textbf {\bibinfo {volume} {329}},\ \bibinfo {pages}
  {659} (\bibinfo {year} {2010})}\BibitemShut {NoStop}%
\bibitem [{\citenamefont {Kuroda}\ \emph {et~al.}(2010)\citenamefont {Kuroda}
  \emph {et~al.}}]{Kuroda2010}%
  \BibitemOpen
  \bibfield  {author} {\bibinfo {author} {\bibfnamefont {K.}~\bibnamefont
  {Kuroda}} \emph {et~al.},\ }\href@noop {} {\bibfield  {journal} {\bibinfo
  {journal} {Phys. Rev. Lett.}\ }\textbf {\bibinfo {volume} {105}},\ \bibinfo
  {pages} {076802} (\bibinfo {year} {2010})}\BibitemShut {NoStop}%
\bibitem [{\citenamefont {Hashimoto}\ \emph {et~al.}(2010)\citenamefont
  {Hashimoto} \emph {et~al.}}]{Hashimoto2010}%
  \BibitemOpen
  \bibfield  {author} {\bibinfo {author} {\bibfnamefont {M.}~\bibnamefont
  {Hashimoto}} \emph {et~al.},\ }\href@noop {} {\bibfield  {journal} {\bibinfo
  {journal} {Nature Phys.}\ }\textbf {\bibinfo {volume} {6}},\ \bibinfo {pages}
  {414} (\bibinfo {year} {2010})}\BibitemShut {NoStop}%
\bibitem [{\citenamefont {Moritz}\ \emph {et~al.}(2011)\citenamefont {Moritz}
  \emph {et~al.}}]{Moritz2011}%
  \BibitemOpen
  \bibfield  {author} {\bibinfo {author} {\bibfnamefont {B.}~\bibnamefont
  {Moritz}} \emph {et~al.},\ }\href@noop {} {\bibfield  {journal} {\bibinfo
  {journal} {Phys. Rev. B}\ }\textbf {\bibinfo {volume} {84}},\ \bibinfo
  {pages} {235114} (\bibinfo {year} {2011})}\BibitemShut {NoStop}%
\end{thebibliography}%

\end{document}